\documentclass[multphys,vecphys]{svmult}

\usepackage{makeidx}         
\usepackage{graphicx}        
\usepackage{multicol}        
\usepackage[bottom]{footmisc}

\makeindex             

\begin{document}

\title*{High-Precision Measurements of $\Delta\alpha/\alpha$
from QSO Absorption Spectra
}
\titlerunning{High-precision $\Delta\alpha/\alpha$ from QSO spectra} 
\author{Sergei A. Levshakov\inst{1}\and
Paolo Molaro\inst{2}\and Sebastian Lopez\inst{3}\and Sandro D'Odorico\inst{4}\and
Miriam Centuri\'on\inst{2}\and Piercarlo Bonifacio\inst{2,5}\and 
Irina I. Agafonova\inst{1}\and Dieter Reimers\inst{6}
}
\authorrunning{Levshakov et al.} 
\institute{Department of Theoretical Astrophysics, Ioffe Physico-Technical
Institute, Politekhnicheskaya Str. 26,  194021 St. Petersburg, Russia
\and  Osservatorio Astronomico di Trieste, Via G. B. Tiepolo 11, 34131 Trieste, Italy 
\and Departamento de Astronom\'ia, Universidad de Chile,
Casilla 36-D, Santiago, Chile
\and European Southern Observatory, Karl-Schwarzschild-Strasse 2,
D-85748 Garching bei M\"unchen, Germany
\and Observatoire de Paris 61, avenue de l'Observatoire, 75014 Paris, France
\and Hamburger Sternwarte, Universit\"at Hamburg,
Gojenbergsweg 112, D-21029 Hamburg, Germany
}
%
%
\maketitle

\begin{abstract}
Precise radial velocity 
\index{precise radial velocity}
measurements 
($\delta v/c \sim 10^{-7}$) of 
Fe\,{\sc ii} lines in damped Ly$\alpha$ systems from
\index{Fe\,{\sc ii} lines}
very high quality VLT/UVES spectra of quasars 
HE 0515--4414 and Q 1101--264
are used to probe cosmological
time dependence of the fine structure constant, $\alpha$. 
\index{fine structure constant}
It is found that between two redshifts $z_1 = 1.15$ and $z_2 = 1.84$
the value of $\Delta \alpha/\alpha$ changes at the level of a few ppm:
$(\alpha_{z_2} - \alpha_{z_1})/\alpha_{\footnotesize 0} = 5.43\pm2.52$ ppm.
Variations of $\alpha$ can be considered as one of the
most reliable method to constrain the dark energy equation of state
\index{dark energy equation of state}
and improvements on the accuracy of the wavelength calibration of
QSO spectra are of great importance.

\end{abstract}

\section{Introduction}
\label{sec_1}

The late-time acceleration of the universe discovered
from the luminosity distance measurements of
high-redshift SNe~Ia
is now regarded as evidence 
for the existence of dark energy.
In many models, the cosmological evolution of dark energy is accompanied by
variations in coupling constants 
as, e.g., the fine-structure constant $\alpha$ \cite{Av}.
Since theories predict different behavior of dark energy, from slow-rolling to
oscillating, to study its dynamics the
coupling constants must be measured with highest
accuracy at each redshift.

In order to fulfill this requirement 
we developed a method called
`{\it Single Ion Differential $\alpha$ Measurement}, SIDAM'
\cite{L04, QRL, L05, L06a}.
Based on the measurements of the relative line position of 
only one element Fe\,{\sc ii}, it is free
from many systematics which affect other methods, e.g., 
\cite{Mu, Ch}. It has been shown that SIDAM can provide a sub-ppm precision
at a single redshift and that this level of accuracy is mainly caused by
systematic errors inherited from the uncertainties of the wavelength scale
calibration \cite{L06a} (ppm stands for parts per million, $10^{-6}$).

In this presentation we consider our recent results \cite{L06b}
with the SIDAM procedure obtained from the analysis of very high
resolution (FWHM $\simeq 3.8$ km~s$^{-1}$, slit width 0.5 arcsec)
and high signal-to-noise (S/N = 100-120 per pixel)
spectra of Q 1101--264 ($z_{\rm em} = 2.15$, $V = 16.0$).
The observations were performed at the VLT Kueyen telescope on 21-23 February, 2006 
under the ESO programme No. 076.A-0463. The total exposure time was 15.4 hours.

\begin{figure}[t!]
\centering
\includegraphics[height=9.5cm,width=11.0cm]{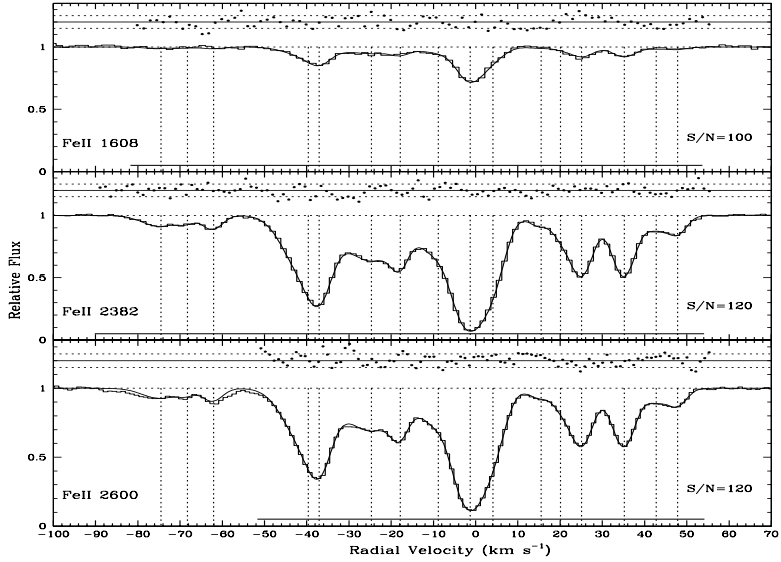}
\vspace{-2.1cm}
\caption{Histograms are the
combined Fe\,{\sc ii} profiles from
the $z = 1.84$ damped Ly$\alpha$ system towards
Q 1101--264 \cite{L06b}.
The zero radial velocity is fixed at $z_{\rm abs} = 1.838911$.
The synthetic profiles are over-plotted by the smooth curves.
The normalized residuals,
$({\cal F}^{cal}_i-{\cal F}^{obs}_i)/\sigma_i$,
are shown by dots (horizontal dotted lines restrict the $1\sigma$ errors).
The dotted vertical lines mark positions of the
sub-components.
Bold horizontal lines mark pixels used to minimize $\chi^2$.
The ranges at $v < -50$ km~s$^{-1}$ and
at $v \simeq -30$ km~s$^{-1}$ in the Fe\,{\sc ii} $\lambda2600$ profile
are blended with weak telluric lines.
}
\label{fig_1}       
\end{figure}

\begin{figure}[t!]
\centering
\includegraphics[height=6.5cm,width=6.5cm]{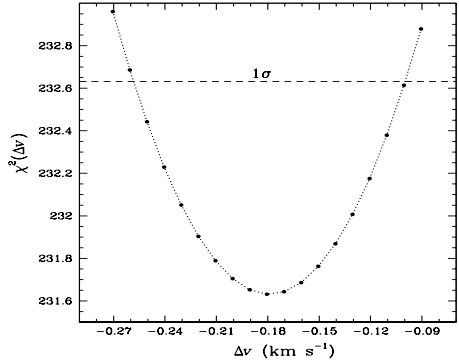}
\vspace{-1.8cm}
\caption{
$\chi^2$ as a function of the velocity difference $\Delta v$
between the Fe\,{\sc ii} $\lambda1608$ and $\lambda2382/2600$
lines \cite{L06b}.
The number of data points $M = 305$,
the number of degrees of freedom $\nu = 257$. 
The minimum of the curve at $\Delta v = -0.18$ km~s$^{-1}$ 
gives the most probable value of 
$\Delta \alpha/\alpha = 5.36$ ppm.
The $1\sigma$ confidence level is determined by $\Delta \chi^2 = 1$
(dashed line) which gives $\sigma_{\Delta v} = 0.08$ km~s$^{-1}$ or
$\sigma_{\Delta \alpha/\alpha} = 2.38$ ppm.
}
\label{fig_2}       
\end{figure}

\section{Results}
\label{sec_2}

The spectroscopic measurability of  
$\Delta \alpha/\alpha \equiv 
(\alpha_z - \alpha_{\footnotesize 0})/\alpha_{\footnotesize 0}$ at
redshift $z$ ($\alpha_{\footnotesize 0}$ refers to the $z = 0$ value) 
is based on the fact that the
energy of each line transition depends individually on a change in
$\alpha$ \cite{W99}.
It means that the relative change of the frequency $\omega_0$
due to varying $\alpha$ is proportional to the so-called
sensitivity coefficient 
\index{sensitivity coefficient}
${\cal Q} = q/\omega_0$ \cite{L05}.
The $q$-values
for the resonance UV transitions in Fe\,{\sc ii} are taken from
\cite{Dz}, and their rest frame wavelengths~--- from \cite{L06b, Ald}. 

The value of $\Delta \alpha/\alpha$ can be 
measured from the relative radial velocity shifts between
lines with different sensitivity coefficients.
In linear approximation ($|\Delta\alpha/\alpha|\ll1$) we have \cite{L06a}:
\begin{equation}
\frac{\Delta\alpha}{\alpha} = \frac{(v_2 - v_1)}
{2\,c\,({\cal Q}_1 - {\cal Q}_2)} =
\frac{\Delta v}{2\,c\,\Delta {\cal Q}}\ ,
\label{E1}
\end{equation}
where index `1' is assigned to the line $\lambda1608$, and
index `2' marks one of the other Fe\,{\sc ii} lines ($\lambda2382$
or $\lambda2600$).

The resulting normalized, vacuum-barycentric, and co-added spectra are
shown in Fig.~\ref{fig_1}.
Two independent data reduction procedures (1D and 2D)
resulted in almost identical co-added spectra.
We found a perfect fit of the Fe\,{\sc ii} profiles to a 
16-component model (shown by the smooth curves in Fig.~\ref{fig_1}):
the normalized $\chi^2$ per degree of freedom equals
$\chi^2_\nu = 0.901$ ($\nu = 257$).
The computational procedure was the same as in \cite{L06a}.
Since the ${\cal Q}$ values for the $\lambda2382$
and $\lambda2600$ lines are equal, their relative velocity shift,
$\Delta v_{2600-2382}$,
characterizes the goodness of wavelength calibrations,
$\sigma_{\rm scale}$, of the 
corresponding echelle orders. 
We found that the value of $\Delta v_{2600-2382} = 20$ m~s$^{-1}$ is
comparable with $\sigma_{\rm scale} \leq 20$ m~s$^{-1}$
estimated from the ThAr lines. So, in what follows
we consider the $\lambda2382$
and $\lambda2600$ lines as having the same radial velocity,
and calculate $\Delta v$ between this velocity
and that of the line $\lambda1608$.

The radial velocity shift between the $\lambda1608$ and $\lambda2382/2600$
lines was derived by comparing synthetic profiles with their observed profiles
and minimizing $\chi^2$.
To find the most probable value of $\Delta v$, we fit the absorption lines
with a fixed $\Delta v$, changing $\Delta v$ in the interval from
$-270$ m~s$^{-1}$ to $-90$ m~s$^{-1}$
in steps of 10 m~s$^{-1}$ (see Fig.~\ref{fig_2}). For each
$\Delta v$, the strengths of the sub-components, their broadening
parameters and relative velocity positions were allowed to vary
in order to
optimize the fit and thus minimize $\chi^2$($\Delta v$).
The most probable value of $\Delta v$ corresponds to the minimum of
the curve $\chi^2$($\Delta v$). 
In our case it is $-180$  m~s$^{-1}$. 
The $1\sigma$ confidence interval
to this value is given
by the condition $\Delta \chi^2 = \chi^2 - \chi^2_{\rm min} = 1$
(the horizontal dashed line in Fig.~\ref{fig_2}.
It is seen from the figure that $\sigma_{\Delta v} = 80$ m~s$^{-1}$, or
$\Delta \alpha/\alpha = 5.36\pm2.38$ ppm.

We have specifically investigated those systematic effects which could 
introduce a non-zero difference between the blue  and 
the red lines ($\lambda1608$ and $\lambda2382/2600$,
respectively) and thus
simulate a variation of $\Delta \alpha/\alpha$ at the ppm level:
(1) possible isotopic shifts caused by unknown isotope abundances,
(2) effects of the unresolved components, and 
(3) possible blends. 
We also compared our previous results \cite{L06b}, where Fe\,{\sc ii} lines
falling in both UVES arms were used, with the present measurements to check up
possible shifts caused by different changes of isophote onto the slit during integration
and did not reveal any of them.
We found that the largest systematic error does not
exceed 30 m~s$^{-1}$, i.e. 
$\sigma_{\Delta\alpha/\alpha,{\rm sys}} \leq 0.89$ ppm\ \cite{L06b}.  
 
We note that the accuracy of 
$\sigma_{\Delta \alpha/\alpha}  = 2.38$ ppm
represents a factor of 1.5 improvement with respect to our 
previous result 
$\sigma_{\Delta \alpha/\alpha}  = 3.8$ ppm  \cite{L05} obtained
from the archive data of lower resolution 
($FWHM \simeq 6$ km~s$^{-1}$) but the same S/N ratio.
Thus the higher spectral resolution 
is shown to significantly contribute to higher
accuracy in the $\Delta \alpha/\alpha$ measurements.

The comparison of
$\Delta \alpha/\alpha = -0.07\pm0.84$ ppm
at $z_1 = 1.15$ towards HE 0515--4414 \cite{L06a} 
with the measured quantity
$\Delta \alpha/\alpha = 5.36\pm2.38$ ppm at $z_2 = 1.84$ 
shows a tentative change of the value of $\Delta \alpha/\alpha$
between these two redshifts:
$(\alpha_{z_2} - \alpha_{z_1})/\alpha_{\footnotesize 0} = 5.43\pm2.52$ ppm.

\bigskip\noindent
{\bf Acknowledgments.}\
SAL acknowledges supports from
the RFBR grant No.~06-02-16489,
the Federal Agency for Science and Innovations
grant NSh~9879.2006.2,
and the DFG project RE 353/48-1.

\printindex
\end{document}